\documentclass[final,5p,times,twocolumn]{elsarticle}
\usepackage{amssymb}
\usepackage{bm}
\usepackage{amsmath}
\usepackage{graphicx}
\usepackage{amsfonts}
\usepackage{ulem}
\usepackage{stmaryrd}
\usepackage{tikz}
\usetikzlibrary{shapes}
\usetikzlibrary{calc}
\usepackage{pgfmath}
\usepackage{pgfplots}
\usepackage{float}
\usepackage{stfloats}
\usepackage{subfig}
\usepackage{natbib}
\usepackage{hyperref}
\usepackage{xcolor}
\pgfplotsset{compat=1.18}

\begin{document}
\begin{frontmatter}
    \title{Electronic origin of solute effects on the mobility of screw dislocation in bcc molybdenum}
    \author[1]{Kangzhi Zhou}
    \author[1]{Jiajun Feng}

        \affiliation[1]{
            organization={School of Physics and Electronics, Hunan Normal University},
%            addressline={No. 15 Taozihu Road, Yuelu District},
            city={Changsha},
            postcode={410081},
            province={Hunan,},
            country={China}
        }

    \author[1,2]{Ziran Liu}
        \ead{zrliu@hunnu.edu.cn}

        \affiliation[2]{
            organization={Key Laboratory of Low-Dimensional Quantum Structures and Quantum Control of Ministry of Education, Key Laboratory for Matter Microstructure and Function of Hunan Province, Department of Physics, Institute of Interdisciplinary Studies, Hunan Normal University},
%           addressline={ADDRESS???},
            city={Changsha},
            postcode={410081},
            province={Hunan,},
            country={China}
        }

            \affiliation[3]{organization={Hunan Provincial Key Laboratory of High-Energy Scale Physics and Applications, School of Physics and Electronics, Hunan University},
%           addressline={ADDRESS???},
            city={Changsha},
            postcode={410082},
            country={China}
            }
             \affiliation[4]{
            organization={Division of Reactor Engineering Technology Research, China Institute of Atomic Energy},
%           addressline={ADDRESS???},
            city={Beijing},
            postcode={102413},
            country={China}
      }

      \author[3]{Huiqiu Deng}
%      \ead{hqdeng@hnu.edu.cn}
      \author[4]{Lixia Jia}
      \author[4]{Xinfu He}
      \ead{hexinfu@ciae.ac.cn}

\begin{abstract}
  In body-centered cubic (bcc) metals such as molybdenum, screw dislocations often exhibit non-Schmid behavior, moving in directions unpredicted by the Schmid law. The mobility of these dislocations is notably influenced by the presence of solute atoms within the alloy matrix. In this study, employing first-principles calculations, we delve into the electronic origins of these influences. Initially, we construct both single atomic column and triple atomic column models to simulate the formation of screw dislocations with solute atoms. Our investigation reveals that tantalum (Ta) and tungsten (W) increase the formation energy of solute-dislocation complexes, in contrast to osmium (Os), iridium (Ir), and platinum (Pt). Subsequently, employing a comprehensive screw dislocation dipole model under shear deformation, we explore the combined effects of solute atoms and deformation on dislocation core movement. Our findings demonstrate that Ta and W, positioned as first nearest neighbors, reduce the stress required to move dislocation cores away from corresponding dislocation dipoles. Conversely, Os, Ir, and Pt exhibit an attractive effect on dislocation cores, lowering the energy barrier for screw dislocation formation and enticing dislocation cores towards these solute atoms.
\end{abstract}

\end{frontmatter}

\section{Introduction}
Screw dislocations play a crucial role in bcc metal as the main carriers of properties such as plastic deformation\cite{proville2012quantum,rodney2017ab,dezerald2016plastic,trinkle2005chemistry}. The presence of solutes in metals can interact with dislocations and strongly modify the nature of the dislocations\cite{trinkle2005chemistry,yu2015origin,leyson2010quantitative}. For example, the solute interacts with the screw dislocation core, which can lead to movement of the dislocation core and thus affect the strength and hardness of the crystal, while the effect of different solute atoms on the screw dislocations varies. Recently, the motion of screw dislocations in bcc metals has been extensively studied by atomistic simulation scales, and the motion of screw dislocations may depend on the influence of several factors, such as solute atoms, deformation or shear stress application \cite{sadananda1972behavior,groger2014stresses,tian2004movement,vitek1976computer,domain2005simulation,gilbert2011stress,woodward2001ab,itakura2012first,pascuet2019solute}.

First, successive descriptions of dislocation dynamics have found that, for each dislocation, there is a stable range of positions, and the actual position of the dislocation depends on its initial position and direction of motion \cite{sadananda1972behavior}. Second, some molecular dynamics or hydrostatic studies have obtained the direction of stress application that moves dislocations, and the critical stress that moves dislocations in pure bcc metals and the slip surface of dislocations Hydrostatic stresses and positive stresses parallel to the line of dislocations do not play a role in dislocation slip \cite{groger2014stresses}. In atomic simulations of W, dislocations do not move until the shear stress increases to a certain value \cite{tian2004movement}. A study utilized three different interatomic interaction potentials to obtain the kinematic properties of screw dislocations\cite{vitek1976computer}. As for the movement of dislocations under stress, the response of a/2$\langle$111$\rangle$  screw dislocation in iron under pure shear strain was studied computationally. The dislocations slip and remain in the (110) plane, and the motion occurs only through nucleation and conduction of double torsion junctions \cite{domain2005simulation,gilbert2011stress}. In addition, in some first-principles calculations, the above statement about the slip plane of the dislocation is corroborated despite the different strains applied. One first-principles study showed that dislocations move on the $\left\{110\right\}$ surface when a stress is applied on the (112) surface in a pure Mo supercell containing 168 atoms \cite{woodward2001ab}. Another study also demonstrated that the migration path of dislocations approaches a straight line and is confined to the $\left\{110\right\}$ plane \cite{itakura2012first}. As for the effect of solute on the dislocation core, experimental studies on the formation and thermal stability of solute clusters near the screw dislocations in FeCuNiMn and FeNiMn alloys have shown that severe segregation of solute atoms prevents the movement of screw dislocations at stresses below 2 GPa \cite{pascuet2019solute}. In conclusion, there is a lack of research on the ease with which solute atoms in Mo affect the formation of dislocation cores and how solute atoms occupying different positions affect the movement of dislocation cores under applied shear deformation.

Research on dislocation dynamics has yielded significant insights into the complex mechanisms governing the motion of dislocations under various conditions. Studies have identified stable ranges of positions for dislocations, with their actual motion dependent on factors such as initial position and direction \cite{sadananda1972behavior}. Additionally, investigations into stress-induced dislocation motion have highlighted the critical role of applied stress direction and magnitude, particularly in pure bcc metals where certain stress orientations are found to facilitate or hinder dislocation slip \cite{groger2014stresses}. Molecular dynamics and hydrostatic studies have revealed that dislocations in materials like tungsten may remain immobile until shear stresses reach specific thresholds. Furthermore, computational simulations have elucidated the kinematic properties of dislocations, showcasing their motion pathways and slip planes, such as the \{110\} surface \cite{itakura2012first}. Despite these advancements, there remains a notable gap in understanding the interplay between solute atoms and dislocation behavior in molybdenum-based systems \cite{pascuet2019solute}. This knowledge deficit underscores the need for further research to elucidate how solute atoms influence dislocation core formation and mobility under applied shear deformation, thereby advancing our comprehension of material behavior at the atomic scale.

Molybdenum, with its typical bcc crystal structure, occupies an important position among the bcc metals and is widely used \cite{yu2016metal,carlen1994concerning,fabritsiev1992effects,walser2007traditional,mannheim2003structural}. In this work, a variety of density functional theory (DFT) approaches are implemented to model screw dislocation generation, structure, and dynamics in bcc Mo with common alloying elements as solutes. Single and triple atomic column arrangements, as well as screw dislocation dipole models, enable examining dislocation core formation energetics and their dependence on solutes Ta, W, Os, Ir, and Pt. Calculation of binding energies quantifies intrinsic solute-dislocation interactions. Furthermore, application of shear deformation reveals the combined effects of solutes and deformation on dislocation core structure and mobility. In fact, we selected the same Mo matrix and transition metal solute atoms as in the previous studies \cite{trinkle2005chemistry,medvedeva2005solid,clouet2009dislocation}. However, we introduced the dipole model and investigated the motion of the dislocation core under shear deformation, which has not been studied before. The comprehensive first-principles analysis provides atomic-scale clarification of solute impacts on fundamental dislocation properties in bcc Mo, with broader implications for plastic deformation mechanisms and alloy development.

\section{Computational methods}
\subsection{First-principles calculations}
The DFT calculations in this work are performed using the Vienna Ab-initio Simulation Package (VASP) \cite{kresse1993ab,kresse1996efficient}, where we use the projector augmented wave (PAW) method \cite{blochl1994projector}, and the exchange correlations are described in the generalized gradient approximation using the Perdew-Burke-Ernzerhof(PBE) function \cite{perdew1996generalized}, with a plane-wave cutoff energy of 291 eV. In the single or triple atomic column model, and in supercells containing dislocation dipoles, the Brillouin zone k-point samples used are 3$\times$3$\times$15 and 1$\times$1$\times$15, respectively \cite{monkhorst1976special}. We obtain the relaxation configurations by the conjugate gradient method and ensure that the force converges to 0.01 eV/$\AA$. It is verified that a more rigorous force convergence criterion does not affect the results. It is worth noting that in the calculations using 36-atom supercells for the single atomic column model and the triple atomic column model, we used static calculations, whereas in the calculations of dislocation formation energies in 135-atom supercells containing complete dislocated dipoles, solute-dislocation interactions, and studies of dislocation cores shifted by the application of shear deformation, for the first two calculations, both the atoms and the supercells were fully relaxed, whereas for the imposed shear deformation calculation, the supercells were fixed and the atoms were allowed to relax only to their most stable state. See Table \ref{table-e-c} for a comparison of the Mo lattice parameters and elastic constants used in this paper and other studies. Our calculation give reasonable lattice parameters and elastic constants when compared with experiment and previous electronic structure calculations.

\begin{table*}[!htb]
\centering
\caption{Lattice parameters and elastic constants of Mo obtained experimentally as well as computationally using different methods or pseudopotentials. FP LMTO: full-potential linear muffin-tin orbital method. USPP: ultrasoft pseudopotentials. PAW: the projector augmented wave (PAW) method.}
\begin{tabular}{cccccc}
\hline
Method & a($\AA$) & K(Mbar) & $\mathrm{C}_{11}(\mathrm{Mbar})$ & $\mathrm{C}_{12}(\mathrm{Mbar})$ & $\mathrm{C}_{44}(\mathrm{Mbar})$ \\
\hline
Experiment\cite{pearson1967handbook} & 3.1468 &  &  &  &  \\
Experiment\cite{simmons1971single} &  & 2.70 & 4.79 & 1.65 & 1.08 \\
FP LMTO\cite{ozolicnvs1993full} & 3.1099 & 2.97 &  &  &  \\
USPP VASP\cite{woodward2001ab} & 3.0996 & 2.88 & 5.05 & 1.79 & 0.92 \\
PAW VASP(this work) & 3.1473 & 2.70 & 4.80 & 1.65 & 0.81 \\
\hline
\end{tabular}\label{table-e-c}
\end{table*}

\subsection{Simulated supercells}
\subsubsection{Single or triple atomic column model}
We selected a triangle formed by three columns of atoms as the simulated dislocation core in a supercell containing 36 Mo atoms and modeled the process of screw dislocation formation by moving the atomic coordinates \cite{medvedeva2005solid} of one or three columns of atoms in the dislocation core along the [111] direction, where the vectors {$\textbf{u}_{1}$,$\textbf{u}_{2}$,$\textbf{u}_{3}$} are defined as: $\textbf{u}_{1}$ = [$\bar{1}$$\bar{1}$2], $\textbf{u}_{2}$ = [1$\bar{1}$0], $\textbf{u}_{3}$ = 1/2[111].  In this supercell, the lattice parameters are ${a}_{1}$ = 15.41$\AA$, ${b}_{1}$ = 13.35$\AA$, and ${c}_{1}$ = 2.72$\AA$ (length of 1\textbf{b}), where $\textbf{a}_{1}$//$\textbf{u}_{1}$,$\textbf{b}_{1}$//$\textbf{u}_{2}$,$\textbf{c}_{1}$//$\textbf{u}_{3}$. The space vectors \textbf{x}, \textbf{y}, and \textbf{z} also remain parallel to $\textbf{u}_{1}$, $\textbf{u}_{2}$, and $\textbf{u}_{3}$, respectively. In this way, the chirality of the dislocation core will complete the transition from clockwise (CW) to counterclockwise (CCW) \cite{trinkle2005chemistry}, while the chirality of the triangle formed by the rest of the surrounding atomic columns remains unchanged. In this case the solute atom is placed in the first nearest neighbor (NN) position of the simulated dislocation core. When modeling the movement of a single column of atoms, we randomly select a column from the three that make up the simulated dislocation core and adjust its fractional coordinates along the [111] direction by $a$ (0 $\leq$ $a$ $\leq$1).

\begin{figure}[h]
	\centering
\includegraphics[width=0.45 \textwidth]{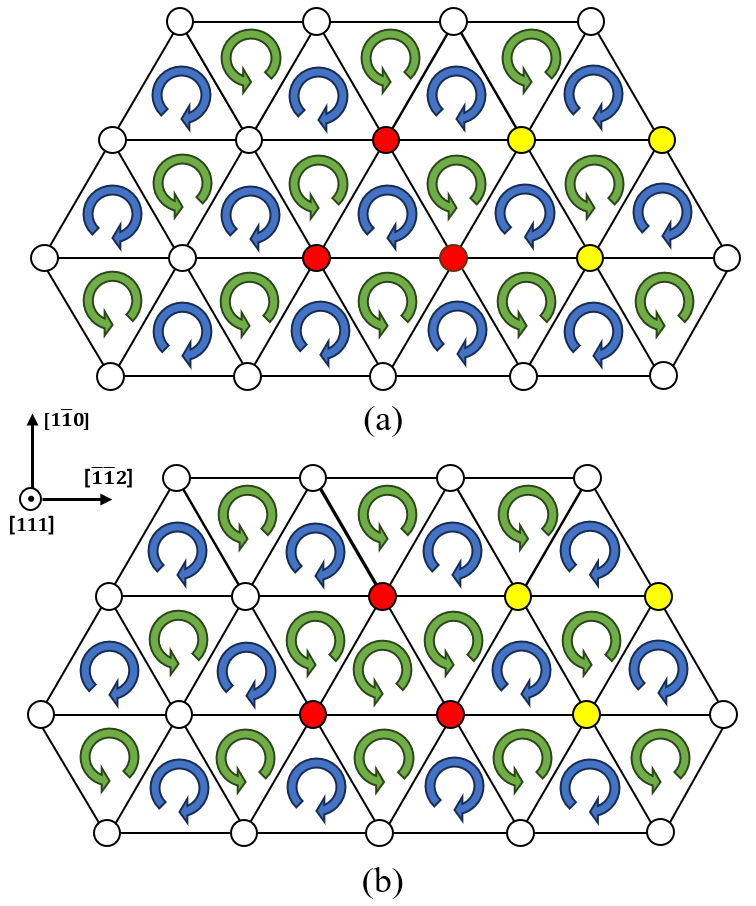}\\
	\caption{The metal atoms are shown as small circles. The atoms that make up the Mo matrix are shown as colorless circles, the Mo atoms that make up the dislocation core are shown in red, and the added positions of the solute atoms are shown in yellow. Clockwise chirality (blue arrows) and counterclockwise chirality (green arrows) are also shown. (a) shows the chirality constituted by the columns of atoms in a perfect crystal, and the change in chirality after moving the columns of atoms is shown in (b).}\label{Figure1}
\end{figure}

Alternatively, in the complete screw dislocation structure, the stable easy core structure in bcc metals is usually represented by a dislocation core consisting of three columns of atoms. In order to model the formation of screw dislocation more accurately, here we perform calculations on the formation of simulated screw dislocation by the motion of three columns of atoms. To model the movement of three columns of atoms, the equation below defines their coordinates along the [111] direction:
\begin{equation}
	x_{1}=a
\end{equation}
\begin{equation}
    x_{2}=1/3+1/3\cdot a
\end{equation}
\begin{equation}
	x_{3}=2/3+2/3\cdot a
\end{equation}
where $x_{1}$ represents the coordinates of the atom moved with initial coordinate 0, $x_{2}$ represents the coordinates of the atom moved with initial coordinate 1/3, and $x_{3}$ represents the coordinates of the atom moved with initial coordinate 2/3 (all along the [111] direction). Here we have calculated the case of solute atoms in three different positions. See Fig. \ref{Figure1} for the variation of the dislocation core in the supercell and the position of the solute atoms.

\subsubsection{Dislocation dipoles and shear deformation application}
\begin{figure}[ht]
	\centering
\includegraphics[width=0.50 \textwidth]{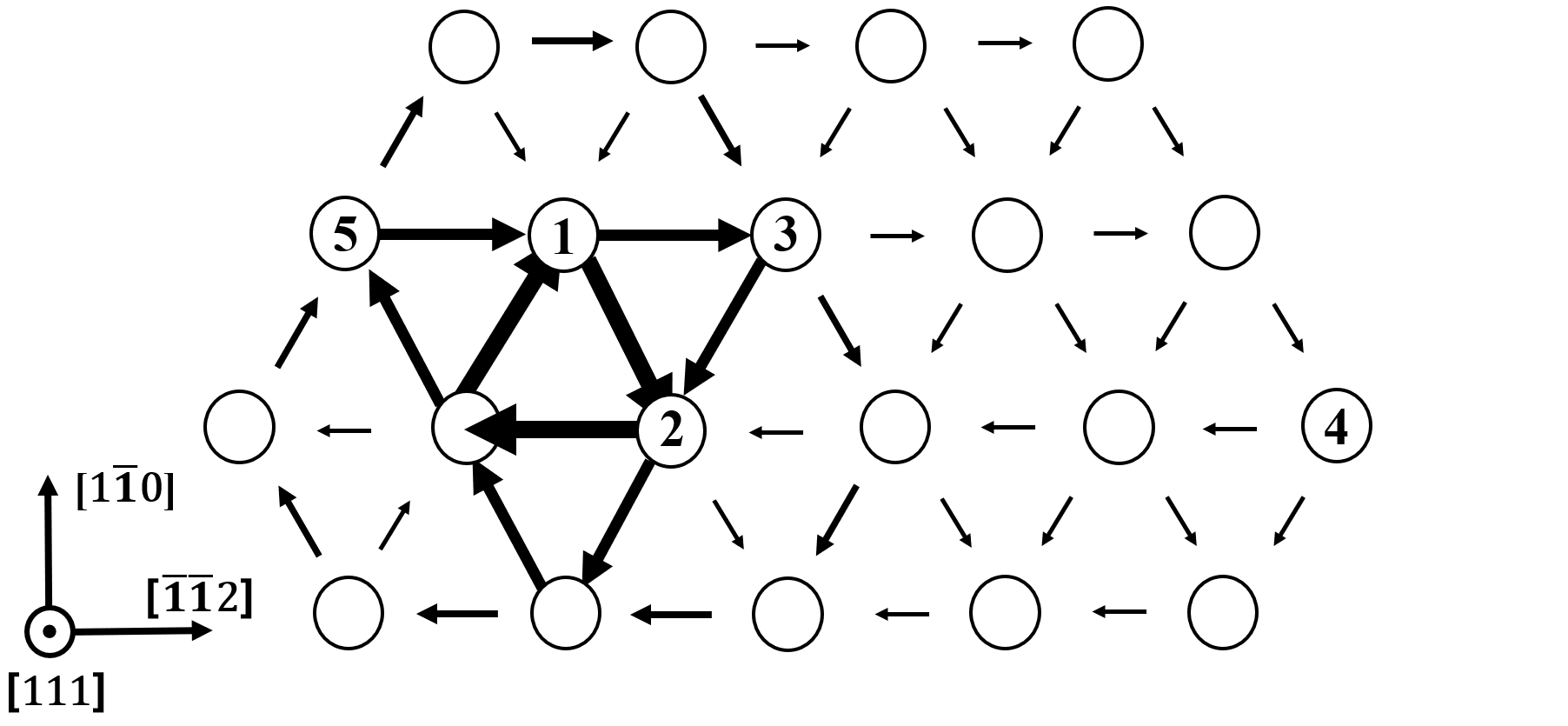}\\
	\caption{Differential displacement map (DDmap) representing the screw dislocation core in the Mo matrix. Numbers 1-5 are the possible locations of solute additions in the study, where 1-2 are inside the dislocation core. The possible positions of solute atoms are 1-3 in the study of calculating the formation energy of dislocations, 1-4 in the study of calculating the interaction of solutes and dislocations, and 3 or 5 in the study of the effect of solute atoms on the movement of dislocation cores under shear deformation.}\label{Figure2}
\end{figure}

We use the Babel package\cite{clouet2011dislocation} to introduce two Burgers vectors of opposite dislocations (135 atoms per Burgers vector) into the supercell by the periodic dislocation dipole method. There are two different dislocation arrangements in the dipole approach. The first is a triangular periodic array of dislocation dipoles, which leads to a triple symmetry of the crystal in the $\langle$111$\rangle$ direction. The second arrangement corresponds to a square periodic array of quadrupoles. The quadrupole arrangement we use converges the energy of the dislocation core more quickly than the triangular arrangement[\cite{ventelon2007core,clouet2009dislocation}]. Here the cell vectors {$\textbf{e}_{1}$,$\textbf{e}_{2}$,$\textbf{e}_{3}$} are defined as: $\textbf{e}_{1}$ = (5/2)$\ast$$\textbf{u}_{1}$ - (9/2)$\ast$$\textbf{u}_{2}$, $\textbf{e}_{2}$ = (5/2)$\ast$$\textbf{u}_{1}$ + (9/2)$\ast$$\textbf{u}_{2}$, $\textbf{e}_{3}$ = $\textbf{u}_{3}$. As discussed previously, this setup was found to be feasible and gives good convergence energies \cite{ventelon2007core,clouet2009dislocation}. Here $\textbf{u}_{1}$, $\textbf{u}_{2}$, and $\textbf{u}_{3}$ are aligned with the above lattice vectors. In this supercell, the lattice parameters are ${a}_{2}$ = 27.96 $\AA$, ${b}_{2}$ = 27.94 $\AA$, and ${c}_{2}$ = 2.72 $\AA$ (length of 1$\textbf{b}$), where $\textbf{a}_{2}$//$\textbf{e}_{1}$,$\textbf{b}_{2}$//$\textbf{e}_{2}$,$\textbf{c}_{2}$//$\textbf{e}_{3}$. Note that the space vectors \textbf{x}, \textbf{y} and \textbf{z} also remain consistent with the space vectors of the 36-atom supercells described above and do not change with the direction of the lattice vectors. See Figures 1 and 2, corresponding to the 36-atom supercell and the 135-atom supercell, respectively. In this case the solute atoms are placed from the dislocation core to the first nearest neighbor (NN) denoted as numbers 1-3. In addition, in calculating the interactions between solutes and dislocations, we place the solute atoms at the positions indicated by the number 4. In the calculation process of applying shear deformation \cite{shimizu2007first,romaner2010effect,li2017impact} to move the dislocation core, we apply the shear deformation by changing the lattice vector. The shear deformation is applied in the direction of $\textbf{e}_{2}$, where the cell vectors {$\textbf{h}_{1}$,$\textbf{h}_{2}$,$\textbf{h}_{3}$} are defined as: $\textbf{h}_{1}$ = $\textbf{e}_{1}$, $\textbf{h}_{2}$ = $\textbf{e}_{2}$ + $x$$\textbf{e}_{3}$ (0 $\leq$ $x$ $\leq$0.25), and $\textbf{h}_{3}$ = $\textbf{e}_{3}$, in which $x$ is the applied deformation. Here $\textbf{e}_{1}$, $\textbf{e}_{2}$, and $\textbf{e}_{3}$ are aligned with the above lattice vectors. As each value of $x$ increases, the atoms are relaxed to a minimum energy state. In the calculation of the applied shear deformation, the solute atoms are placed at the positions indicated by numbers 3 and 5. Please refer to Fig. \ref{Figure2} for details.

\section{Results and discussion}
\subsection{Single atomic column model}
\begin{figure}[h]
	\centering
\includegraphics[width=0.45 \textwidth]{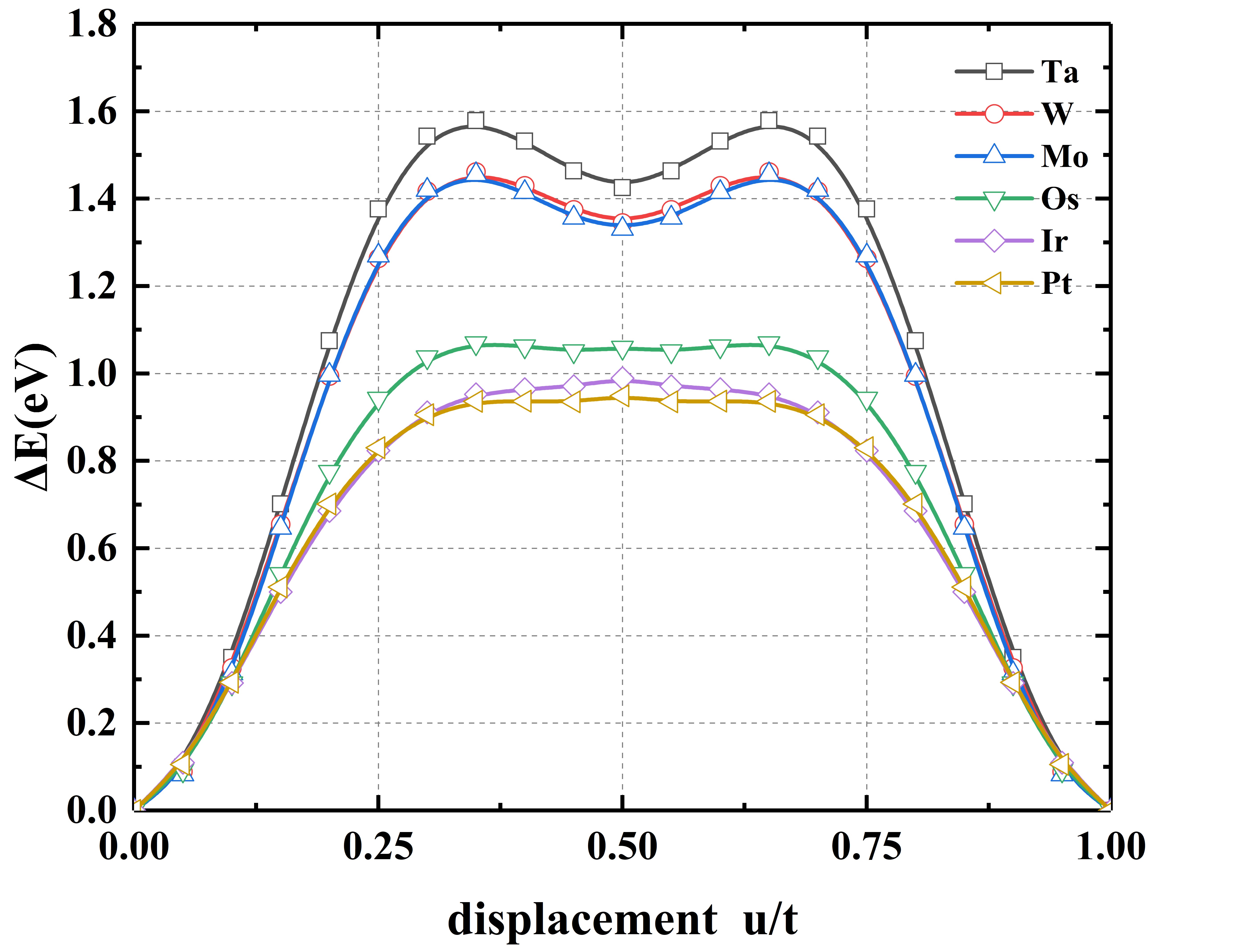}\\
	\caption{Energy difference for Mo and Mo alloys during atomic column movement. Displacement is normalized by the magnitude of the Burgers vector 1/2 $\langle$111$\rangle$.}\label{Figure3}
\end{figure}

In this model, we move only one atomic column to simulate the formation of screw dislocations, which is manifested as a change in the chirality of the three atomic columns selected as dislocation cores, and we simulate the ease of forming screw dislocation in Mo under the influence of solute atoms by placing solute atoms in the first nearest neighbor of the dislocation cores, and the corresponding energy difference have been calculated for the addition of different solute atoms and for different degrees of movement of the atomic columns. Fig. \ref{Figure3} shows that higher energy values are required for the formation of screw dislocation under the influence of Ta and W, while lower energies are required in the presence of Os, Ir, and Pt. In fact, this law is consistent with the hardening or softening effect of solute atoms on Mo. Solutes with lower $d$-electron numbers (Ta and W) harden the alloy, while solutes with higher $d$-electron numbers (Os, Ir and Pt) soften the alloy \cite{medvedeva2005solid,medvedeva2005electronic}.

\subsection{Triple atomic column model}
\begin{figure}[!t]
	\centering
\includegraphics[width=0.45\textwidth]{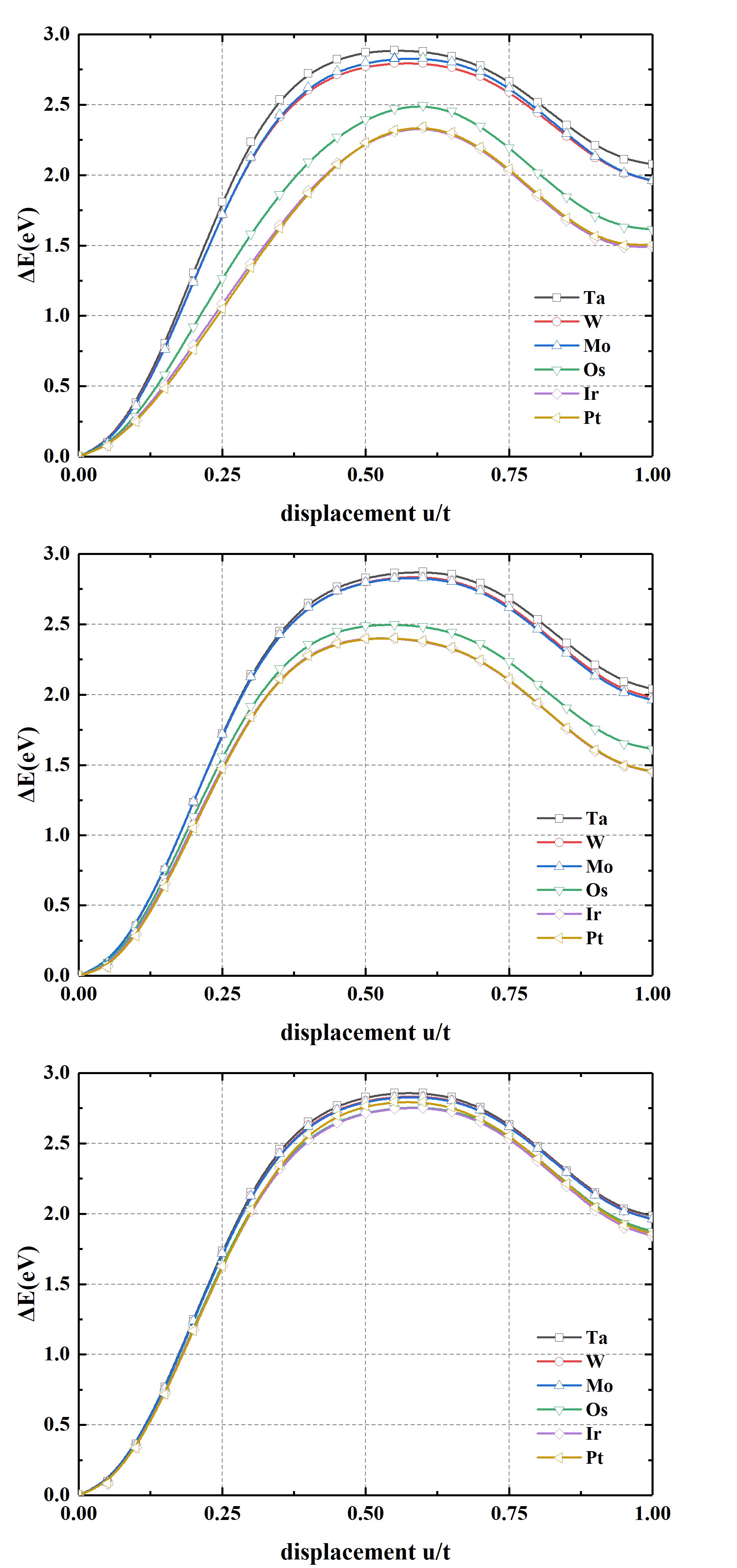}\\
	\caption{Corresponding to the changes in the Energy difference caused by shifting the atomic columns when the solute atom is in the first nearest neighbor (NN), second nearest neighbor (2NN), and third nearest neighbor (3NN) of a dislocation core consisting of three atomic columns, respectively.}\label{Figure4}
\end{figure}

In this model, we place the solute atoms in three different positions to investigate the effect of the solute atom positions on the simulated dislocation cores, e.g. whether the different positions have an effect on the laws of the ease of forming screw dislocation in Mo under the influence of solute atoms, and to compare them with the dislocation dipoles that will be investigated below. From Fig. \ref{Figure4} we can see that the  law remain consistent with the single atomic column model, and their effects on the dislocation core gradually converge as the solute atoms move away from the dislocation core.

\subsection{Dislocation dipole formation energy}
\begin{figure}[h]
	\centering
\includegraphics[width=0.45 \textwidth]{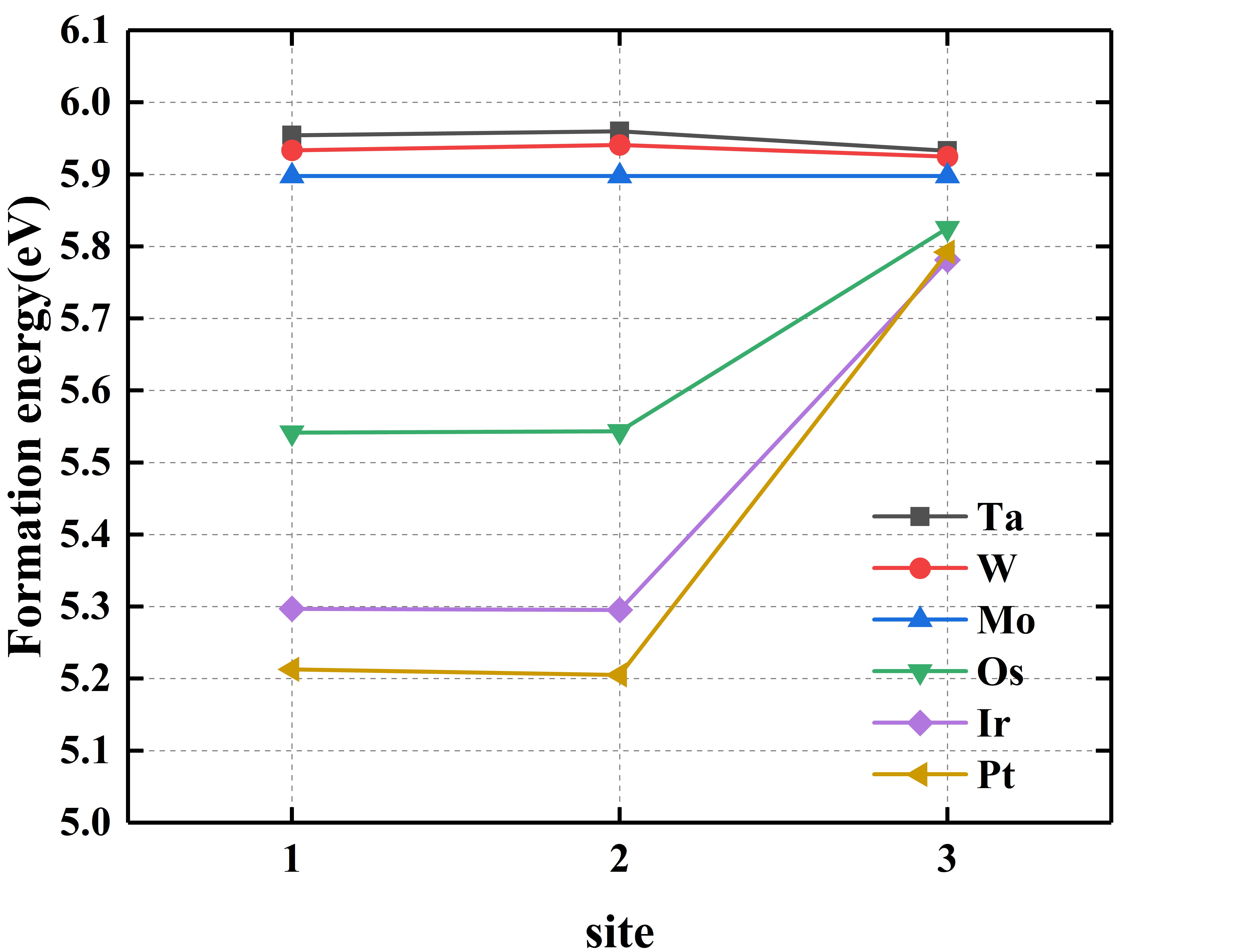}\\
	\caption{Effect of different solutes on the formation energy of dislocations when placed at different positions. Site1 and site2 are inside the dislocation core and site3 is at the first nearest neighbor (NN) of the dislocation core.Refer to Figure \ref{Figure2} for the positions of the solute atoms for the horizontal coordinates.}\label{Figure5}
\end{figure}

We use supercells with dislocation dipoles as end states and perfect or supercells with solute atoms (without dislocation cores) as initial states to model the process of dislocation core formation and calculate the corresponding formation energies from the energy difference between them. From Fig. \ref{Figure5}, the law of difficulty of screw dislocation formation induced by solute atoms is consistent with that obtained in the single or triple atomic column model. Meanwhile, the formation energy curves obtained for Ta and W when they are placed as solute atoms from site1 to site 3 are smoother, while the formation energies of Os, Ir, and Pt undergo a larger shift from the inside of the dislocation core to the first nearest neighbor of the dislocation core, and the difference between individual solutes also decreases. We hypothesize that this is still related to the number of $d$-electrons in the solutes.

\subsection{Solute-dislocation interactions}
\begin{figure}[h]
	\centering
\includegraphics[width=0.45 \textwidth]{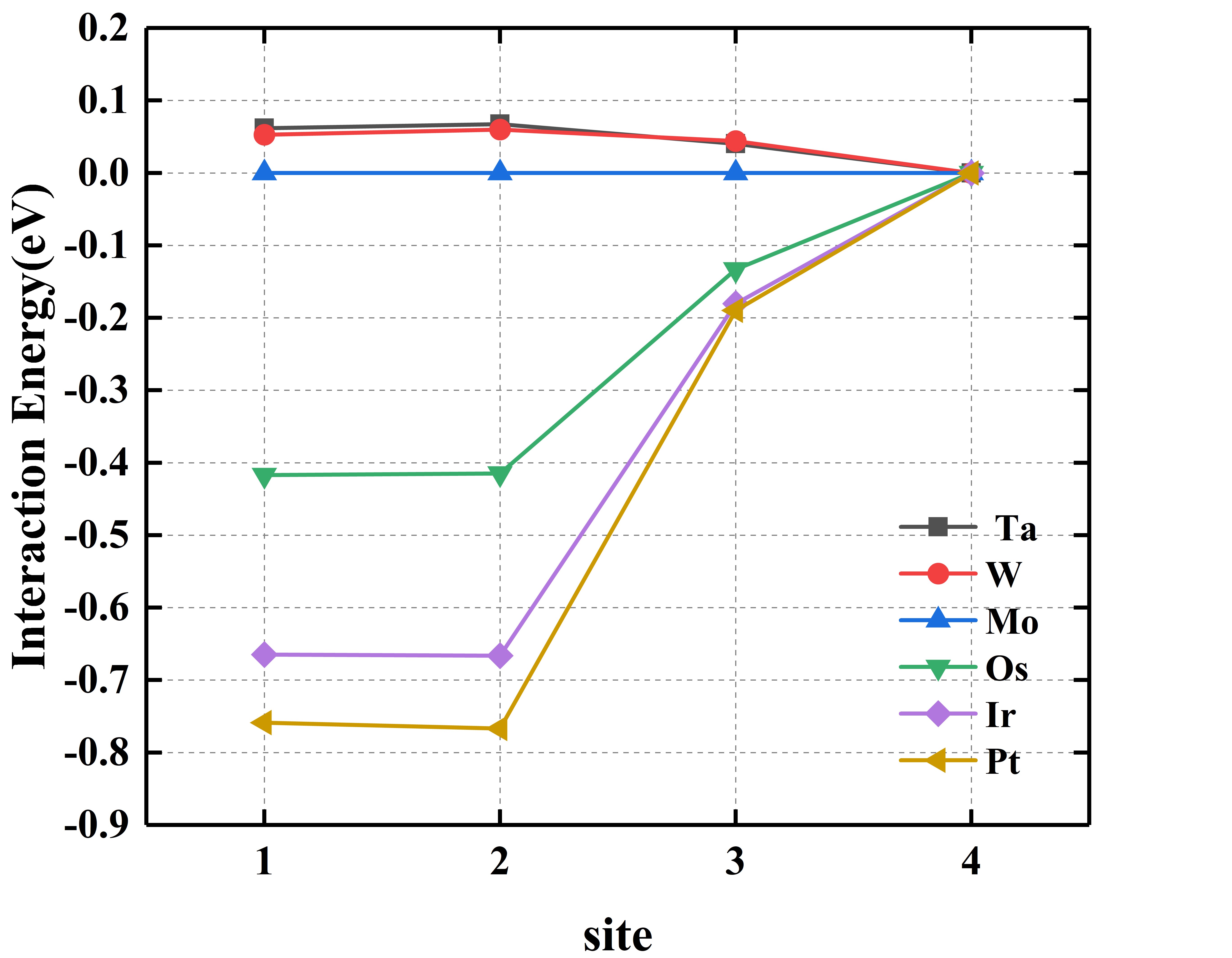}\\
	\caption{Interaction energy between solutes and screw dislocations.Ta and W exhibit repulsion and Os, Ir and Pt exhibit attraction. Refer to Figure \ref{Figure2} for the positions of the solute atoms for the horizontal coordinates.}\label{Figure6}
\end{figure}

Here we have calculated the energy of a solute atom placed at a position far away from the dislocation (number 4 in the Fig. \ref{Figure2}) as the initial state, and calculated the energy difference between the solute at positions 1-3 and the initial state to simulate the movement of a solute atom from a distance toward the dislocation core. In the Fig. \ref{Figure6} we can see that Ta and W have repulsive effects on the dislocation core, while Os, Ir and Pt have attractive effects. This law is also consistent with the previously mentioned law of ease of dislocation core formation, i.e., it is difficult to form screw dislocations in the vicinity of solute atoms that have repulsive interactions, and vice versa.

 \begin{figure}[!t]
 	\centering
 	\subfloat[]{
 			\includegraphics[scale=0.33]{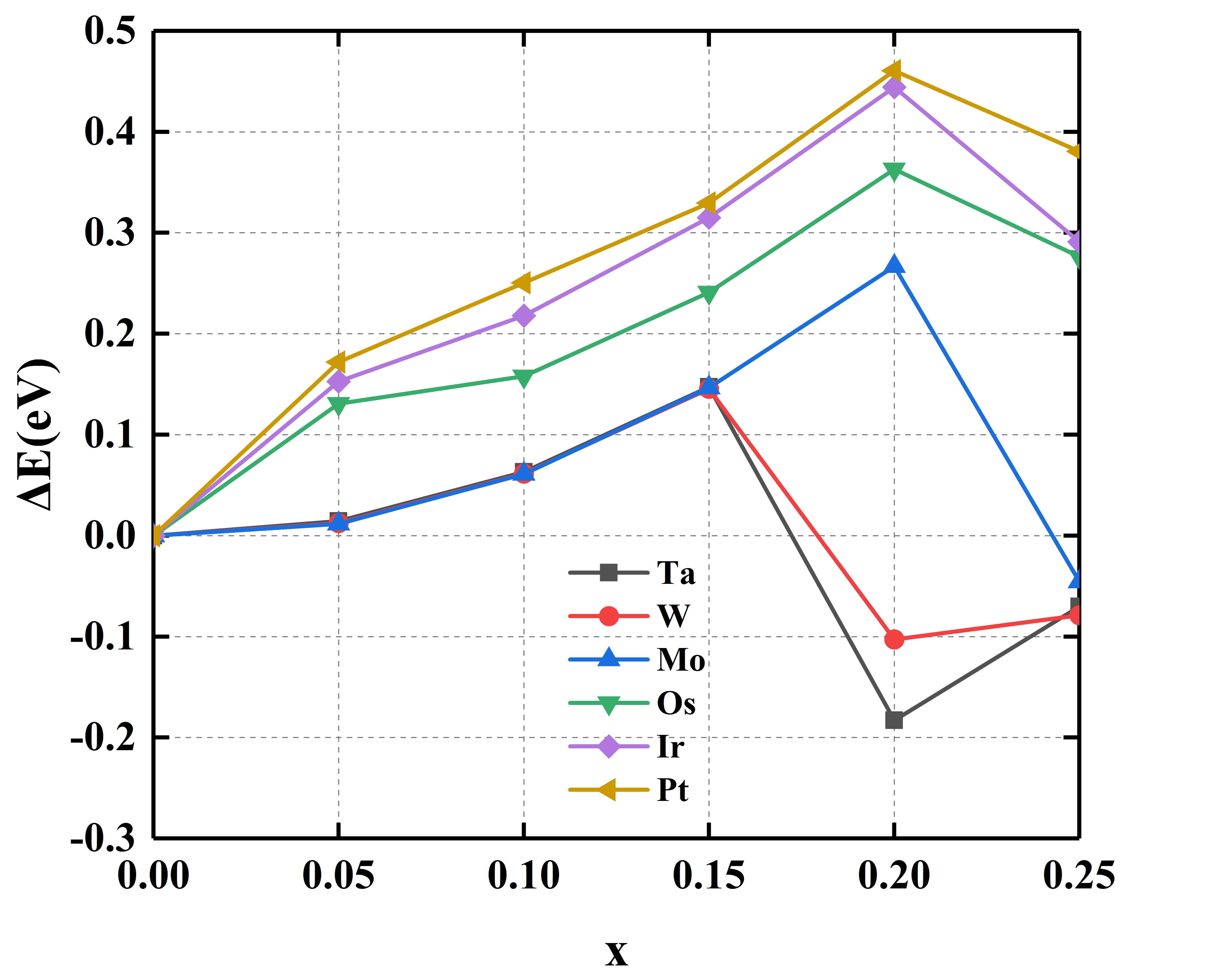}}\\
 	\subfloat[]{
 			\includegraphics[scale=0.3]{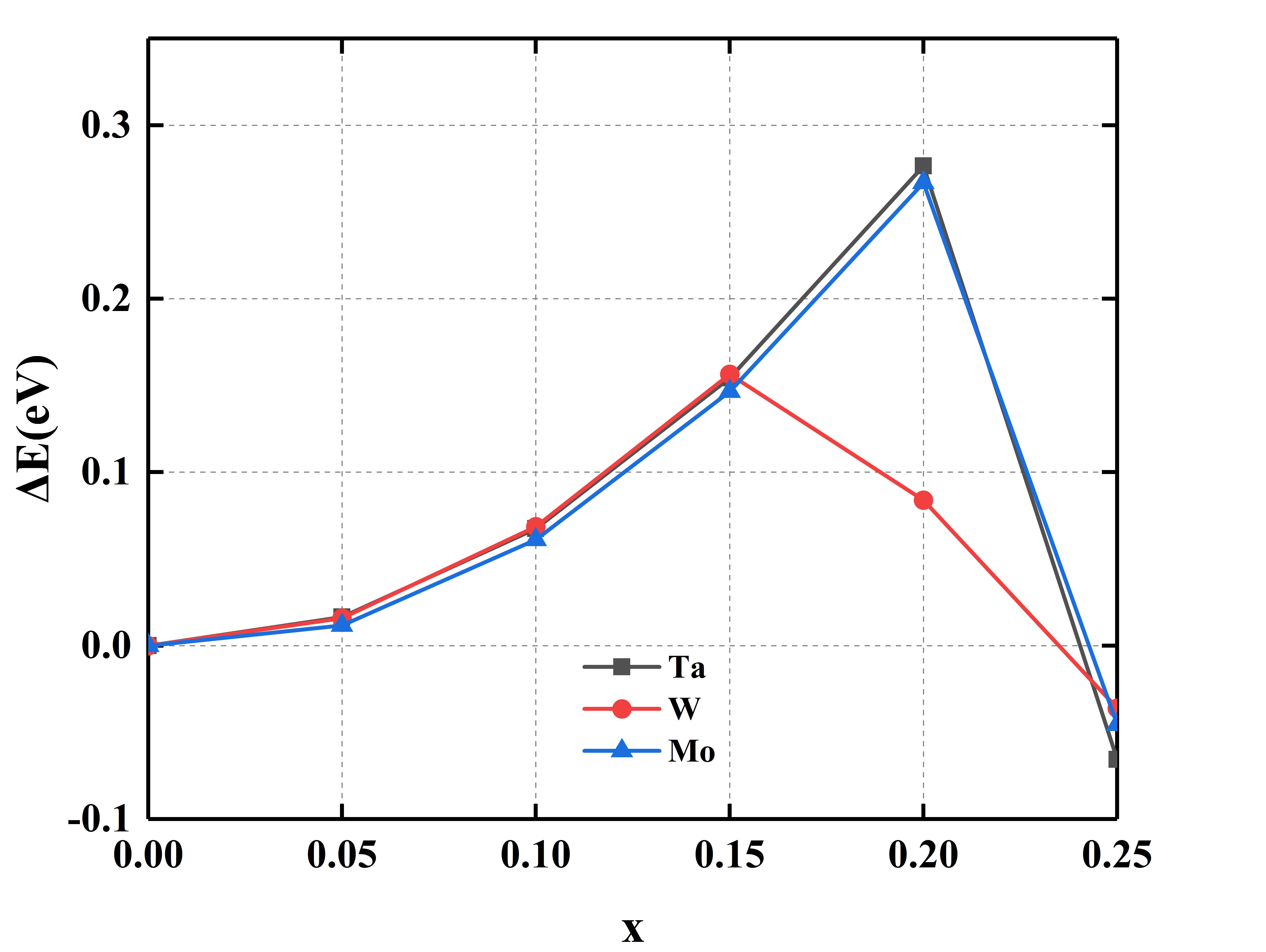}}
 	\caption{The change in the difference between the energy of the supercell and the energy of the supercell before deformation is applied as the shear deformation increases. In (a)(b), the solute atoms are in different first nearest neighbors of the dislocation core. The solute atom is in position 3 in (a) and in position 5 in (b). Refer to Figure \ref{Figure2} for the positions of the solute atoms. Refer to 2.2.2 for the definition of horizontal coordinates.}\label{Figure7}
 \end{figure}

\begin{figure}[!t]
	\centering
	\subfloat[]{
		\includegraphics[scale=0.2]{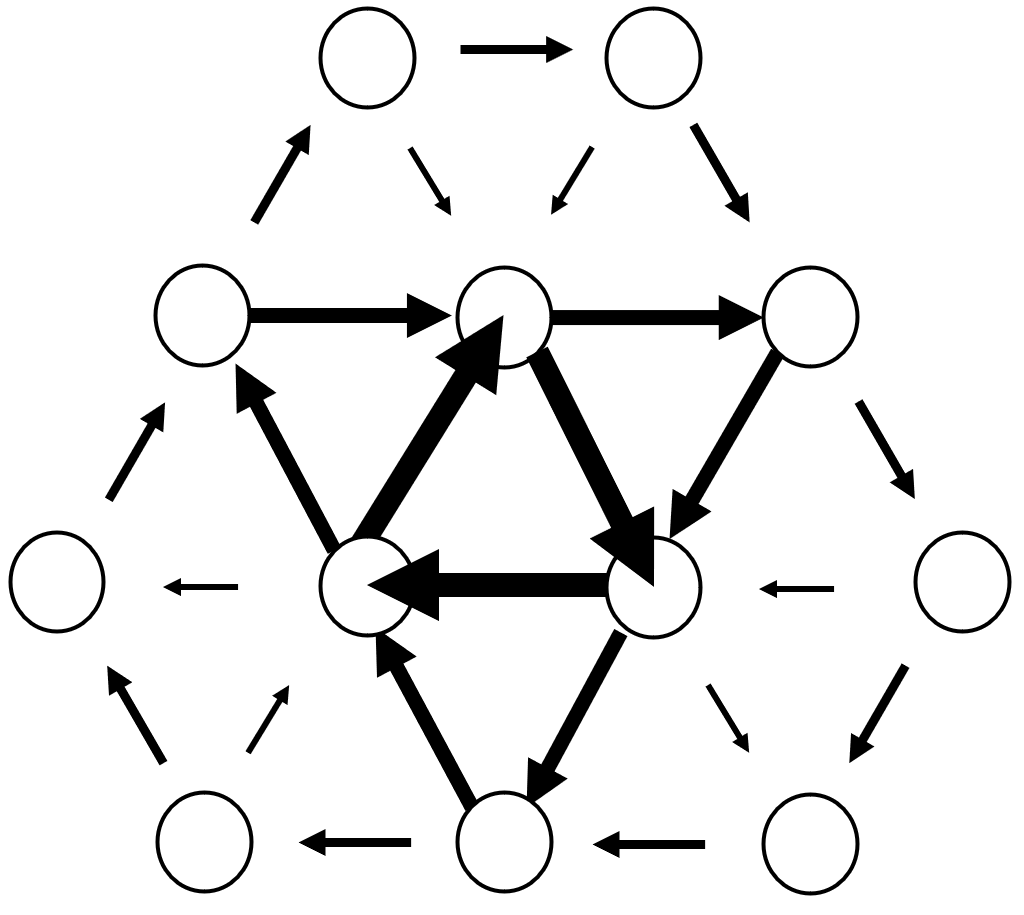}}
	\subfloat[]{
		\includegraphics[scale=0.2]{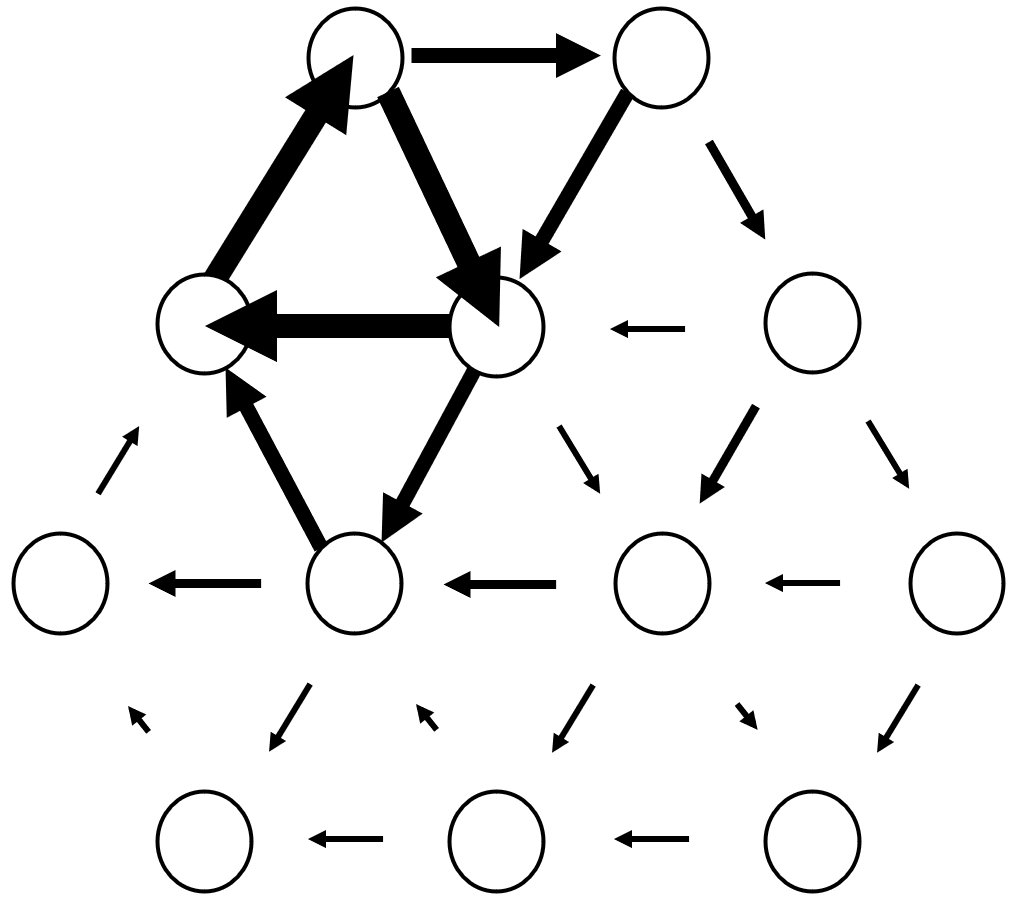}}
	\\
	\subfloat[]{
		\includegraphics[scale=0.2]{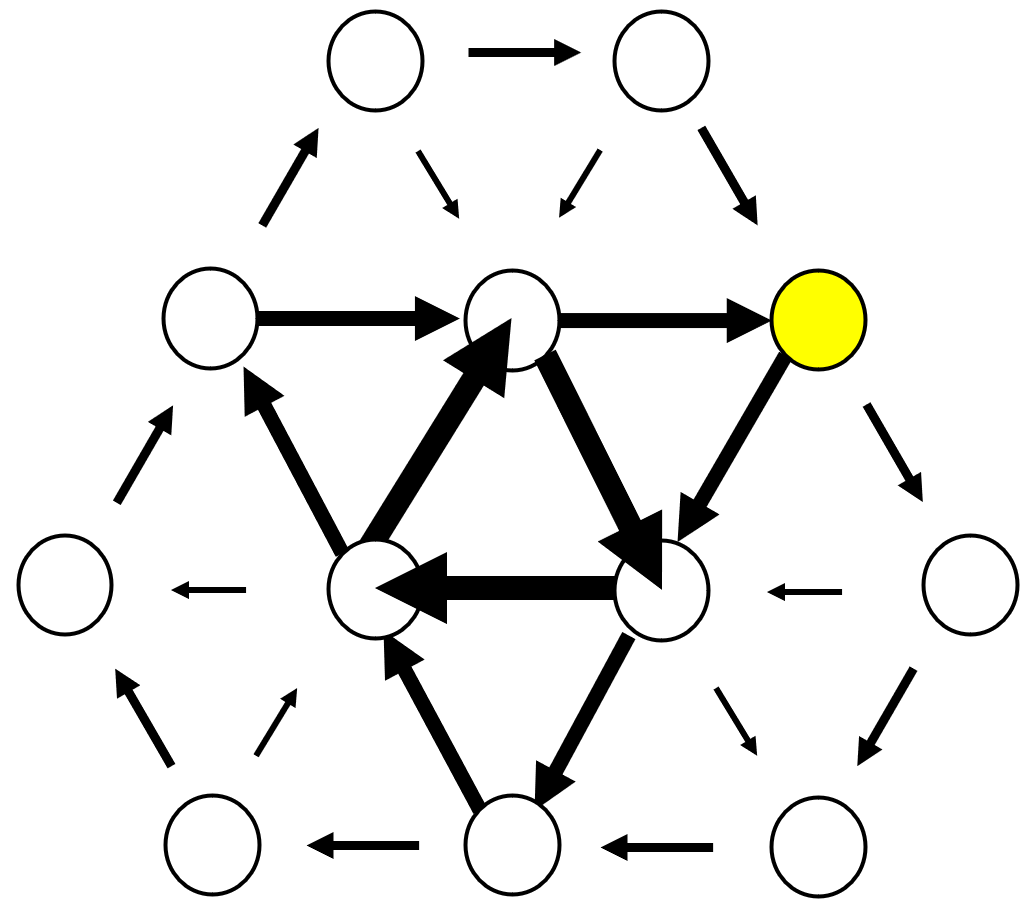}}
	\subfloat[]{
		\includegraphics[scale=0.2]{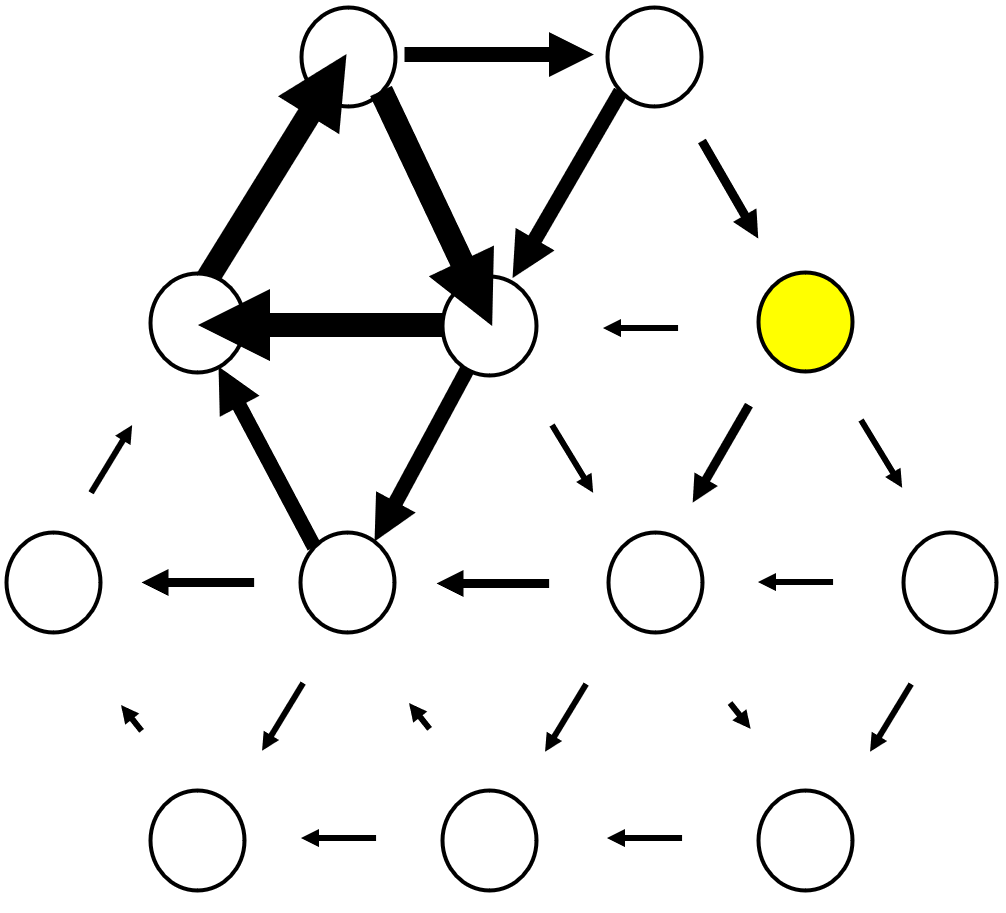}}
	\\
	\subfloat[]{
		\includegraphics[scale=0.2]{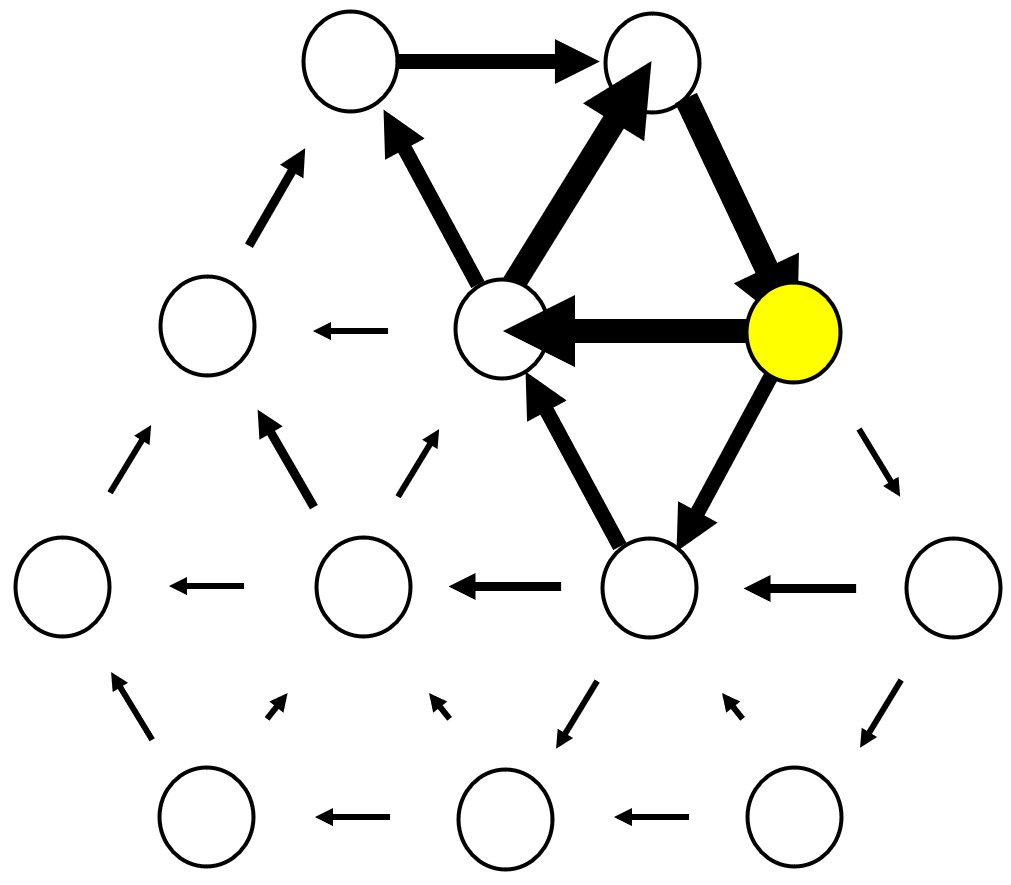}}
	\\
	\subfloat[]{
		\includegraphics[scale=0.2]{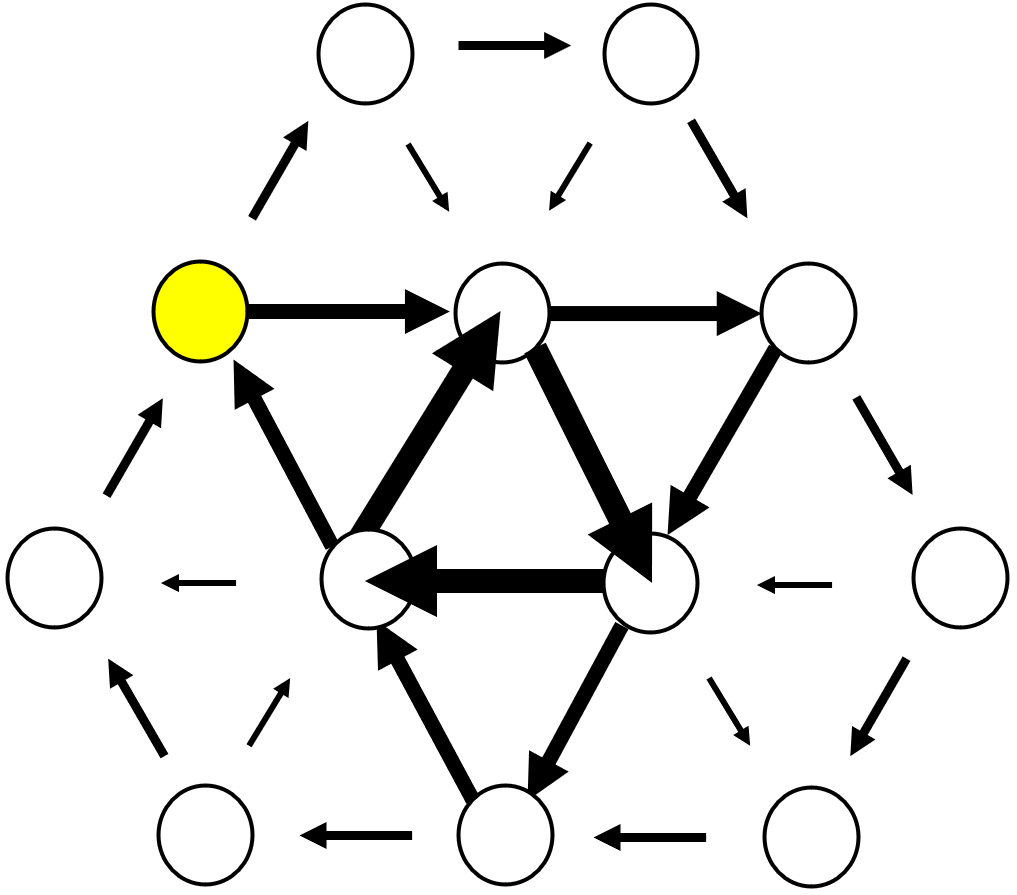}}
	\subfloat[]{
		\includegraphics[scale=0.2]{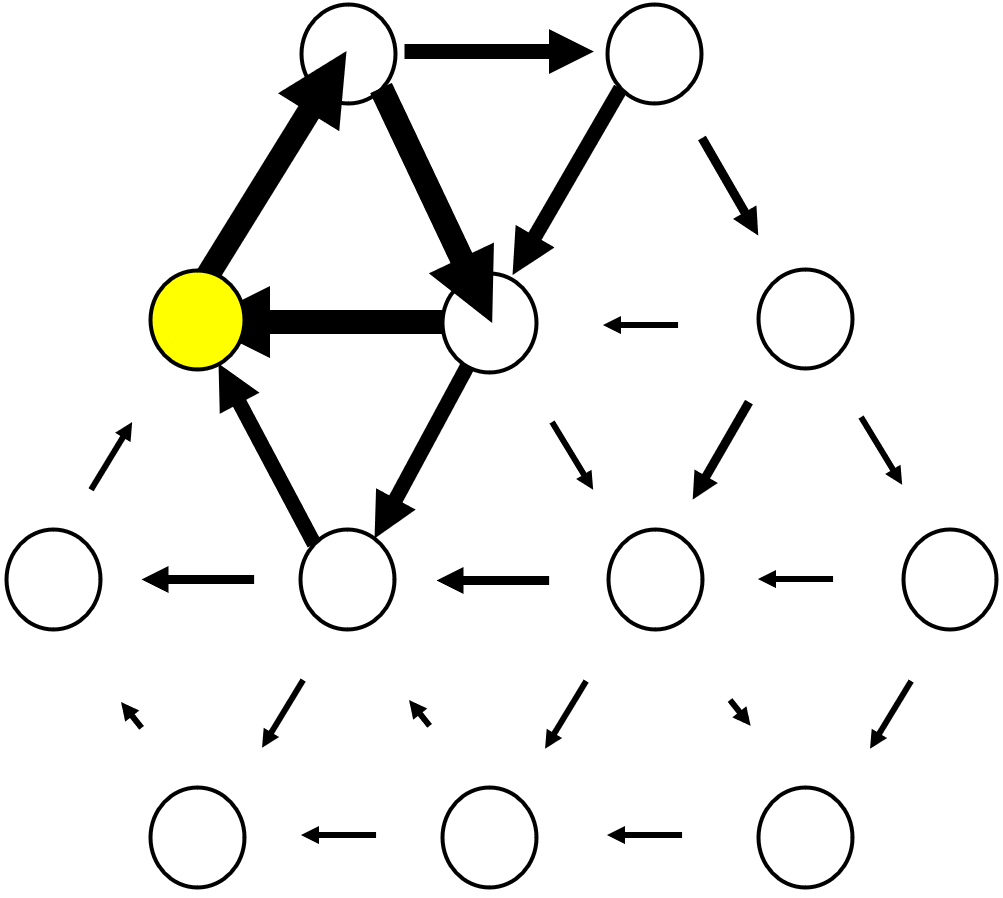}}
	\\
	\caption{(a) (b) Representation of the initial state of pure Mo before the application of shear deformation and the position of the dislocation core after the movement has occurred, respectively. Correspondingly, (c)(f) represent the initial state of the solute atoms when they are at the positions numbered 3 and 5 shown in Fig. 2. (d)(e) represent the positions of the dislocation core after the application of sufficiently large deformation when the atoms with repulsive effect on the dislocation core (Ta, W) and the atoms with attractive effect (Os, Ir, Pt) are at the position numbered 3, respectively. (g) is the state of the dislocation core after the application of sufficient deformation when the W atom is in the number 5 position.}
	\label{Figure8}
 \end{figure}

\subsection{Shear deformation application}
We have applied shear deformation to a supercell containing a dislocation dipole of 135 atoms to observe the movement of the dislocation core, and also plotted the energy change of the supercell with increasing deformation as shown in Fig. \ref{Figure7}. We find that the dislocation core moves in different directions in both the \{110\} and \{112\} planes with the addition of deformation and different solute atoms, and that this movement differs with the increase in deformation and the type and position of solute atoms. It is worth noting that when the solute atoms are outside the dislocation core, i.e., at positions 3 and 5 shown in Figure \ref{Figure2}, the dislocation core may remain in its original position or move toward the solute atoms under the influence of the applied deformation and the interaction of the solute atoms. While positions 1 and 2 shown in Figure \ref{Figure2} are internal to the dislocation core, applying the same amount of deformation when the solute atoms are in positions 1 and 2 does not cause the dislocation core to move away from the solute atoms in our calculations, and thus no valid conclusions can be drawn. Therefore, in this section of the study of applied shear deformation, the solute atoms are selected at positions 3 and 5 as shown in Figure\ref{Figure2}.

Here we take the dislocation core on the side of the added solute atom as an example and illustrate its movement using a differential displacement map (DDmap). We first discuss the solute atoms at number 3 shown in Fig. \ref{Figure2}. First, for pure Mo, Fig. \ref{Figure8} (a)(b) shows the initial state of the dislocation core in Mo with no solute atoms and no deformation applied, and the position of the dislocation core after deformation is applied to move the dislocation, which corresponds to the state of the bent Mo in Fig. \ref{Figure7} at x = 0 and x = 0.25. The dislocation core does not move during the increase of x from 0 to 0.20, when the energy of the system increases due to the application of deformation. x = 0.25 is the critical value, and the dislocation core moves in the direction away from the corresponding dislocation dipole to release the stress, and the energy decreases and is lower than that of the initial state shown in (a), where the dislocation core is in its original position. This indicates that the dislocation releases stress by moving away from the corresponding dislocation dipole. (c) (d) (e) shows how the solute atoms affect the motion of the dislocation core. (c) shows the initial state where the solute atom is located in the first immediate neighbor of the dislocation core but has not yet applied any deformation, and (d) shows the motion of the dislocation core under the influence of Ta and W. We find that the motion remains consistent with pure Mo, but from Fig. \ref{Figure7} we can see that the dislocation core has moved at x = 0.2, i.e. with a smaller deformation compared to pure Mo. This is due to the repulsive effect of Ta and W on the dislocation core. Correspondingly, the motion of the dislocation core is altered under the influence of the attracting solute atoms. (e) shows the motion of the dislocation core under the influence of Os, Ir and Pt. Due to the attraction of the solute atoms to the dislocation core, the dislocation core moves towards the solute atoms, i.e. the solute atoms form a new dislocation core with the two surrounding atomic columns, and the effect of this interaction on the dislocation core is greater than that of the shear deformation. As for x = 0.25, the decrease in energy of the supercell with Os, Ir and Pt added is due to the fact that the dislocation dipole on the side of the unadded solute atom also moves. This result is in good agreement with the ease of dislocation core formation under the influence of solute atoms and the interaction between solute atoms and dislocation cores calculated in the previous work.

For the solute atom at position 5 in Fig. \ref{Figure2}, if the solute atom has an attractive effect on the dislocation core, i.e., Os, Ir, and Pt, the effect of the solute atom's attraction to the dislocation core and the effect of the movement of the dislocation core due to the application of deformation (referring to Fig. \ref{Figure8}(a)(b)) would overlap. We will only study the effects of atoms that have repulsive effects on the dislocation core. Fig. \ref{Figure8} (f) shows the initial state, while (g) shows the state in which the dislocation core moves after sufficient deformation is applied when the W atom is in the number 5 position. It is worth noting that Ta do not move the dislocation core under the influence of Ta atoms at x = 0.25 in this position, whereas the dislocation core moves towards the solute atoms both in pure Mo and with added W atoms. At this point, in the supercell with added W atoms, the motion effect of the deformation on the dislocation core is greater than the repulsion effect between the W atoms and the dislocation core. This also confirms that the repulsive effect of Ta is stronger than that of W. And for the energy decrease in the supercell with W added at x = 0.20, it is also due to the movement of the dislocation dipole on the side of the unadded solute atom.

\section{Conclusion}
In conclusion, our study employs first-principles calculations to systematically analyze the interaction energies between solute atoms and screw dislocations in molybdenum (Mo). We constructed a single atomic column model, a triple atomic column model, and a complete dislocation dipole for the study. For the general conclusions that can be drawn from them, solute atoms with a known hardening effect on Mo, such as Ta and W, exhibit a repulsive impact on dislocation cores, leading to an increased energy requirement for screw dislocation formation. The investigation of combined effects of solute atoms and deformation on dislocation core movement reveals that Ta and W in the first nearest neighbors reduce the stress necessary to move dislocation cores away from corresponding dislocation dipoles. Conversely, Os, Ir, and Pt demonstrate an attractive effect on dislocation cores, resulting in a decreased energy barrier for screw dislocation formation and a tendency for dislocation cores to be drawn toward these solute atoms. Furthermore, when solute atoms are positioned as first nearest neighbors on the opposite side, it is confirmed that the addition of Ta and W impedes the movement of dislocation cores, with Ta exerting a stronger hindering effect than W. Among their differences, the previous two atomic models for simulating screw dislocations are unable to study the dislocation core under the joint influence of deformation and solute atoms by applying deformation, whereas the dipole method can obtain the slip surface and direction of the dislocation core under the joint influence of deformation and solute atoms compared to the previous two methods. In addition, in terms of computational cost, the first two methods use static calculations to obtain the order of energy barrier heights corresponding to the order of magnitude of the interactions between solute atoms and dislocation cores, whereas the dipole method, which establishes a dipole and undergoes a complete relaxation to calculate the interactions, is much more costly to compute. In other words, if only the interactions of different solute atoms with respect to the dislocation core need to be compared, it is sufficient to use either the single atomic column model or the triple atomic column model. Overall, our individual calculations are in strong agreement and offer valuable predictive insights into the intricate ways in which solute atoms influence the formation and movement of dislocation cores, particularly at low concentrations.

\section*{Declaration of Competing Interest}
The authors declare that they have no known competing financial interests or personal relationships that could have influenced the work reported in this paper.

\section*{Acknowledgments}
The authors are grateful for financial support from National Key Research and Development Program of China (2023YFB3002100), National Natural Science Foundation of China (U2341260, 51671086) and Natural Science Foundation of Hunan Province (2022JJ30032).

\bibliographystyle{elsarticle-num-names}
\bibliography{reference.bib}
\end{document}